\documentclass{aastex}
\usepackage{spr-astr-addons}
\usepackage{url}

\RequirePackage{color}

\begin{document}

\title{Dynamics of interacting phantom and quintessence dark
energies}

\shorttitle{Dynamics of interacting dark energies}
\shortauthors{Farooq et al.}

\author{M. Umar Farooq\altaffilmark{1}}
\and
\author{Mubasher Jamil\altaffilmark{2}}
\and
\author{Ujjal Debnath\altaffilmark{3}}

\altaffiltext{1}{Center for Advanced Mathematics and Physics,
National University of Sciences and Technology, H-12, Islamabad,
Pakistan. Email: m\_ufarooq@yahoo.com}

\altaffiltext{2}{Center for Advanced Mathematics and Physics,
National University of Sciences and Technology, H-12, Islamabad,
Pakistan. Email: mjamil@camp.nust.edu.pk , jamil.camp@gmail.com}

\altaffiltext{3}{Department of Mathematics, Bengal Engineering and
Science University, Shibpur, Howrah-711 103, India. Email:
ujjaldebnath@yahoo.com , ujjal@iucaa.ernet.in}

\begin{abstract}
We present models, in which phantom energy interacts with two
different types of dark energies including variable modified
Chaplygin gas (VMCG) and new modified Chaplygin gas (NMCG). We then
construct potentials for these cases. It has been shown that the
potential of the phantom field decreases from a higher value with
the evolution of the Universe.
\end{abstract}


\section{Introduction}

One of the outstanding developments in cosmological physics in the
past decade is the discovery of the accelerated expansion of the
universe,
supposedly driven by some exotic dark energy \citep{perl,ries,sper,sper1,cope}.
 Surprisingly, the energy density of the dark energy is two-third
of the
critical density ($\Omega_\Lambda\simeq0.7$) apart from dark matter ($%
\Omega_m\simeq0.3$). The astrophysical data shows that this sudden
transition in the expansion history of the universe is marginally recent ($%
z\simeq0.7$) compared with the age of the universe. The nature and
composition of dark energy is still an open problem. With the
thermodynamical studies of dark energy, it is conjectured that the
constituents of dark energy may be massless particles (bosons or
fermions) whose collective behavior resembles a kind of radiation
fluid with negative pressure. Moreover, the temperature of the
universe filled with dark energy will increase as the universe
expands \citep{lima}. The earliest proposal to
explain the recent accelerated expansion was the cosmological constant $%
\Lambda$ represented by the equation of state (EoS) $p=-\rho$ (or
$w=-1$) having a negative pressure. In order to comply with the
data, the cosmological constant has to be fine tuned up to 120
orders of magnitude \citep{doglov}, which requires extreme fine
tuning of several cosmological parameters. The cosmological constant
also poses a famous cosmic coincidence problem (the question of
explaining why the vacuum energy came to dominate the universe very
recently) \citep{bento1}. The coincidence problem is
tackled with the use of a homogeneous and time dependent scalar field $\phi$,
 in which the scalar field rolls down a potential $V(Q)$ according
to an attractor-like solution to the equations of motion
\citep{zlat}. But here the field has difficulties in reaching
$w<-0.7$, while current observations favor $w<-0.78$ with 95\%
confidence level \citep{linder}. Other scalar field models of dark
energy include ghost condensates \citep{ghost}, tachyon
\citep{sen,setare2}, holographic dark energy \citep{holo} and quintom \citep%
{quintom}. It is shown that a quintessence scalar field coupled with
either a dissipative matter field, a Chaplygin gas (CG) or a
tachyonic fluid solves the coincidence problem \citep{chim}. These
problems are alternatively discussed using anthropic principles as
well \citep{wein}. Several other models have been proposed to
explain the cosmic accelerated expansion by introducing decaying
vacuum energy \citep{free,frie}, a cardassian term in the
Friedmann-Robertson-Walker (FRW) equations \citep{free1}, a
generalized Chaplygin gas (GCG) \citep{bento,setare1,setare3} and a
phantom energy ($w<-1$) arising from the violation of energy
conditions \citep{cald,babi,ness,setare}. Another possibility is the
`geometric dark energy' based on the Ricci scalar $R$ represented by
$\Re=R/12H^2$, where $H$ is the Hubble parameter \citep{linder}.
Notice that $\Re>1/2$ represents accelerated expansion, and $\Re>1$
gives a super-accelerated expansion of the universe,
whereas presently $\Re=1/2$.\\

Models based on dark energy interacting with dark matter have been
widely investigated
\citep{vage,sami,lin,wu,wang1,jamil,jamil1,zim,seta,rub,mota}. These
models yield stable scaling solution of the FRW equations at late
times of the evolving universe. Moreover, the interacting CG allows
the universe to cross the phantom divide (the transition from $w>-1$
to $w<-1$), which is not permissible in pure CG models. In fact it
is pointed out that a phantom divide (or crossing) is possible only
if the cosmic fluids have some interaction \citep{vik}. It is
possible that this interaction can arise from the time variation of
the mass of dark matter particles \citep{zhang1}. It is shown that
the cosmic coincidence problem is fairly alleviated in the
interacting CG models \citep{campo}. This result has been endorsed
with interacting dark energy in \citep{sad}. There is a report that
this interaction is physically observed in the Abell cluster A586,
which in fact
supports the GCG cosmological model and apparently rules out the $\Lambda$%
CDM model \citep{bertolami}. However, a different investigation of
the observational $H(z)$ data rules out the occurrence of any such
interaction and favors the possibility of either more exotic
couplings or no interaction at all \citep{wei,umar,umar1}. The
consideration of interaction between quintessence and phantom dark
energies can be motivated from the quintom models \citep{xin}. In
this context, we have investigated the interaction of the dark
energy with dark matter by using a more general interaction term. We
have focused on the inhomogeneous EoS for dark energy as these are
phenomenologically relevant.\\

The outline of the paper is as follows. In the section II, we
present a general interacting model for our dynamical system.
Following \citep{Chatto}, we consider the two interacting dark energy
models like variable modified Chaplygin gas (VMCG) and new modified
Chaplygin gas (NMCG) interact with phantom field in sections III and
IV. We found the phantom potential in these scenarios. Finally, we
present our conclusion.

\section{The model}

We assume the background to be a spatially flat isotropic and
homogeneous
FRW spacetime, given by%
\begin{equation}
ds^{2}=dt^{2}-a^{2}(t)[dr^{2}+r^{2}(d\theta ^{2}+\sin ^{2}\theta
d\phi ^{2})],
\end{equation}%
where $a(t)$ is the scale factor. The corresponding Einstein field
equations
are%
\begin{equation}
3H^{2}=\rho _{\text{tot}}
\end{equation}%
and
\begin{equation}
6(\dot{H}+H^{2})=-(\rho _{\text{tot}}+3p_{\text{tot}}).
\end{equation}%
Here $\rho _{\text{tot}}$ and $p_{\text{tot}}$ represent the total
energy density and isotropic pressure respectively ($8\pi G=c=1$).
Moreover, the
energy conservation for our gravitational system is given by%
\begin{equation}
\dot{\rho}_{\text{tot}}+3H(\rho _{\text{tot}}+p_{\text{tot}})=0.
\end{equation}%
Suppose we have a two-component model of the form%
\begin{equation}
\rho _{\text{tot}}=\rho _{1}+\rho _{2},
\end{equation}%
and%
\begin{equation}
p_{\text{tot}}=p_{1}+p_{2}.
\end{equation}%
Here $\rho _{1}$ and $p_{1}$ denote the energy density and pressure
of quintessence and $\rho_{2}$, $p_{2}$ denote the energy density
and pressure of phantom dark energy. The stress energy tensor for
matter-energy is
\begin{equation}
T_{\mu \upsilon }=-\partial_\mu \Phi \partial _{\upsilon }\Phi
-g_{\mu \upsilon }\Big[\frac{\sigma }{2}g^{\beta \delta }\partial
_{\beta }\partial _{\delta }\Phi +V(\Phi )\Big].
\end{equation}

By assuming that the phantom field is evolving in an isotropic
homogenous universe and that $\Phi$ is merely function of time, from
Eq. (7) one
 can extract energy density and pressure as
\begin{eqnarray}
\rho _{1} &=&\frac{\sigma }{2}\dot{\Phi}^{2}+V(\Phi ), \\
p_{1} &=&\frac{\sigma }{2}\dot{\Phi}^{2}-V(\Phi ).
\end{eqnarray}
Here $\sigma=-1$ corresponds to the phantom field while $\sigma=+1$
represents the standard scalar field which represents the
quintessence field, also $V(\Phi)$ is the potential. In this case,
the equation of state $w$ is given by
\begin{equation}
w=\frac{p_1}{\rho_1}=\frac{\sigma\dot\Phi^2-2V(\Phi)} {\sigma\dot%
\Phi^2+2V(\Phi)}.
\end{equation}
We observe that it results in the violation of the null energy condition $%
\rho_1+p_1=\sigma\dot{\Phi}^2>0$, if $\sigma=-1$. Since the null
energy condition is the basic condition, its violation yields other
standard energy
conditions to be violated likewise dominant energy condition ($\rho_1>0$, $%
\rho_1\geq|p_1|$) and the strong energy condition ($\rho_1+p_1>0 $, $%
\rho_1+3p_1>0 $). Due to the energy condition violations, it makes
the failure of cosmic censorship conjecture and theorems related to
black hole thermodynamics. The prime motivation to introduce this
weird concept in cosmology does not come from the theory but from
the observational data. According to the forms of dark energy
density and pressure (8) and (9), one can easily obtain the kinetic
energy and the scalar potential terms as
\begin{eqnarray}
\dot{\Phi}^2&=&\frac{1}{\sigma}(1+\omega)\rho_1, \\
V(\Phi)&=&\frac{1}{2}(1-\omega)\rho_1.
\end{eqnarray}

\section{Variable Modified Chaplygin Gas}

Firstly, let us suppose that we have variable modified Chaplygin
gas (VMCG) representing the dark energy and is given by \citep{UDeb}%
\begin{equation}
p_{2}=A_{1}\rho _{2}-\frac{B_{0}a(t)^{-n}}{\rho _{2}^{\alpha }},
\end{equation}%
where $0\leq \alpha \leq 1,0\leq A_{1}\leq 1,$ $B_{0}$ and $n$ are
constant parameters. The Chaplygin gas behaves like dust in the
early evolution of the universe and subsequently grows to an
asymptotic cosmological constant at late time when the universe is
sufficiently large. In the cosmological context, the Chaplygin gas
was first suggested as an alternative to quintessence
\citep{kamenshchik}. Later on, the Chaplygin gas state equation was
extended to a modified form by adding a barotropic term
\citep{Benaoum,debnath2,jamil123}. Recent supernovae data also favor
the two-fluid cosmological model with Chaplygin gas and matter
\citep{grigoris}. Suppose that the phantom field interacts with
(VMCG), so under this interaction (supposing the interaction term is
$Q$) the continuity equations can be written as
\begin{eqnarray}
\dot{\rho}_{1}+3H(\rho _{1}+p_{1}) &=&Q, \\
\dot{\rho}_{2}+3H(\rho _{2}+p_{2}) &=&-Q.
\end{eqnarray}
In case of $Q=0,$ we arrive at the non-interacting situation while
$Q>0$ exhibit a transfer of energy from the one fluid of density
$\rho _{1}$ to other fluid of density $\rho _{2}.$ In order to solve
the above continuity equations different forms of $Q$ have been
considered. Here we will proceed to solve the continuity equation
(15) by taking $Q=3\delta H\rho _{2}$ ($\delta$ is a coupling
constant), so we get
\begin{eqnarray}
\rho _{2}&=&\Big[ \frac{3B_{0}(1+\alpha )}{[3A_{1}(1+\alpha
)+3(1+\alpha
)(1+\delta )-n]a^{n}}\nonumber\\&&+\frac{C}{a^{3(1+\alpha )(1+\delta +A_{1})}}\Big] ^{%
\frac{1}{1+\alpha }},
\end{eqnarray}
where $C$ is the constant of integration. One can be seen that if
$n=0$ and $A, B$ approache to zero, then $\rho _{2}\sim
a^{-3(1+\delta )}$. Now for simplicity, we choose
$V=m\dot{\Phi}^{2}$, where $m$ is a positive constant. So using (14)
and (16), we obtain the kinetic term as
\begin{eqnarray}
\dot{\Phi}^{2}&=&C_{1}a^{-\frac{6\sigma}{\sigma+2m}}+\frac{6\sigma(1+\alpha)}{(-2mn+(6-n+6\alpha)\sigma)}
\nonumber\\&&\times\Big(
\frac{3B_{0}(1+\alpha)}{-n+3(1+\alpha)(1+\delta+A_{1})}
\Big)^{\frac{1}{1+\alpha}}~a^{-\frac{n}{1+\alpha}}
\nonumber\\&&\times_{2}F_{1}\Big[x,-\frac{1}{1+\alpha},1+x,
-Ya^{n-3(1+\alpha)(1+\delta+A_{1})}\Big],\nonumber\\
\end{eqnarray}
and the potential energy has the form
\begin{eqnarray}
V&=&mC_{1}a^{-\frac{6\sigma}{\sigma+2m}}+\frac{6m\sigma(1+\alpha)}{(-2mn+(6-n+6\alpha)\sigma)}\nonumber\\&&\times\Big(
\frac{3B_{0}(1+\alpha)}{-n+3(1+\alpha)(1+\delta+A_{1})}
\Big)^{\frac{1}{1+\alpha}}~a^{-\frac{n}{1+\alpha}}
\nonumber\\&&\times_{2}F_{1}\Big[x,-\frac{1}{1+\alpha},1+x,
-Ya^{n-3(1+\alpha)(1+\delta+A_{1})}\Big],\nonumber\\
\end{eqnarray}
where
\begin{eqnarray*}
Y&=&\frac{C(-n+3(1+\alpha)(1+\delta+A_{1}))} {3B_{0}(1+\alpha)},\\
x&=&\frac{2mn+(n-6(1+\alpha))\sigma}{(1+\alpha)(2m+\sigma)(-n+3(1+\alpha)
(1+\delta+A_{1}))},
\end{eqnarray*}
and  $C_{1}$ is the constant of integration. From the expression
(18) it is clear that the potential energy is a function of scale
factor $a$. The graphs represented by Fig. 1 and Fig. 2 show that
$\phi$ increases with the passage of time
while the $V$ decreases with the increase of cosmic time $t$.\\

\section{New Modified Chaplygin Gas}

The model which behaves as a dark matter (radiation) at the early
stage and \textit{X-}type dark energy at late stage is the New
Modified Chaplygin Gas
(NMCG) \citep{Zhang2,Chatto1}%
\begin{equation}
p_{2}=\beta \rho _{2}+\frac{wA_{2}a^{-3(1+w)(1+\alpha )}}{\rho
_{2}^{\alpha }},\ \ A_{2}>0,\ \ \beta >0.
\end{equation}%
In view of $($second energy eq.$),$ the energy density of the (NMCG)
can be
expressed as%
\begin{equation}
\rho_{2} =\left[ \frac{A_{2}wa^{-3(1+w)(1+\alpha )}}{w-\delta -\beta
}+C_{1}a^{-3(1+\delta +\beta )(1+\alpha )}\right]
^{\frac{1}{1+\alpha }}
\end{equation}%
Now for simplicity, we again choose $V=m\dot{\Phi}^{2}$, where $m$
is a positive constant. So using (14) and (20), we obtain the
kinetic term as
\begin{eqnarray}
\dot{\Phi}^{2}&=&C_{2}a^{-\frac{6\sigma}{\sigma+2m}}-\frac{2a^{-3(1+w)}\sigma}{(2m(1+w)+\sigma(-1+w))}
\nonumber\\&&\times
_{2}F_{1}\Big[x_1,-\frac{1}{1+\alpha},1+x_1,Y_1a^{3(1+\alpha)(w-\beta-\delta)}
\Big],\nonumber\\
\end{eqnarray}
and the potential energy has the form
\begin{eqnarray}
V&=&mC_{2}a^{-\frac{6\sigma}{\sigma+2m}}-\frac{2ma^{-3(1+w)}\sigma}{(2m(1+w)+\sigma(-1+w))}
\nonumber\\&&\times
{2}F_{1}\Big[x_1,-\frac{1}{1+\alpha},1+x_1,Y_1a^{3(1+\alpha)(w-\beta-\delta)}
\Big],\nonumber\\
\end{eqnarray}
where
\begin{eqnarray*}
Y_1&=&\frac{C_{1}(-w+\beta+\delta)}{A_{2}w{2}},\\
x_1&=&\frac{-2m(1+w)+\sigma(1-w)}{(1+\alpha)(2m+\sigma)(w-\beta-\delta)},
\end{eqnarray*}
and $C_{2}$ is the constant of integration.

\section{Discussion}

In this work, we have considered the interacting scenario of the
universe, in which phantom energy interacts with two different types
of dark energies including variable modified Chaplygin gas (VMCG),
new modified Chaplygin gas (NMCG). By considering some particular
form of interaction term, we have constructed the potential of the
phantom field. By looking at the energy conservation equations (14)
and (15) it is found that the energies of the (VMCG) and (NMCG) are
getting transferred to the phantom field. With the help of graphs we
studied the variations of $V$ and $\phi$ with the variation of the
cosmic time. From the figures we see that the potential decreases
from the lower value with the evolution of the universe. Thus in the
presence of an interaction, the potential decreases and the field
decreases with the evolution of the universe.

\begin{figure}
\includegraphics[height=2in]{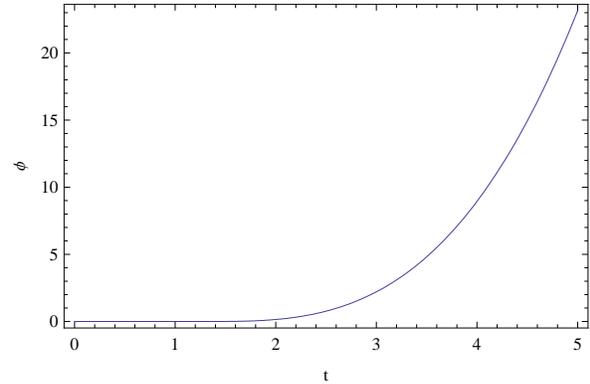}\\
\caption{The variation of
$\Phi$ against cosmic time $t$ with particular values of parameters
$A_{1}=0.1,B_{0}=0.1,\delta=0.005,\alpha=0.5,C=1,C_{1}=1,n=0.9,m=0.7$.}
\end{figure}
\begin{figure}
\includegraphics[height=2in]{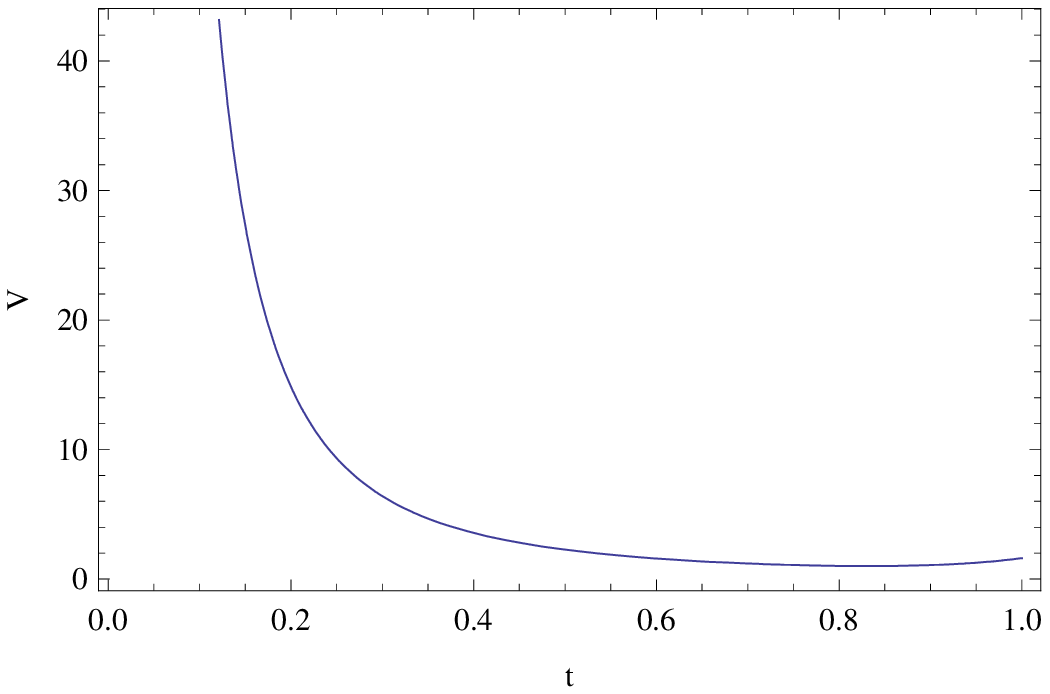}\\
\caption{The variation of
$V$ against cosmic time $t$ $\sigma=-1$
(phantom field) in VMCG with particular values of parameters
$A_{1}=0.1,B_{0}=0.1,\delta=0.005,\alpha=0.5,C=1,C_{1}=1,n=0.9,m=0.7$.}
\end{figure}
\begin{figure}
\includegraphics[height=2.5in]{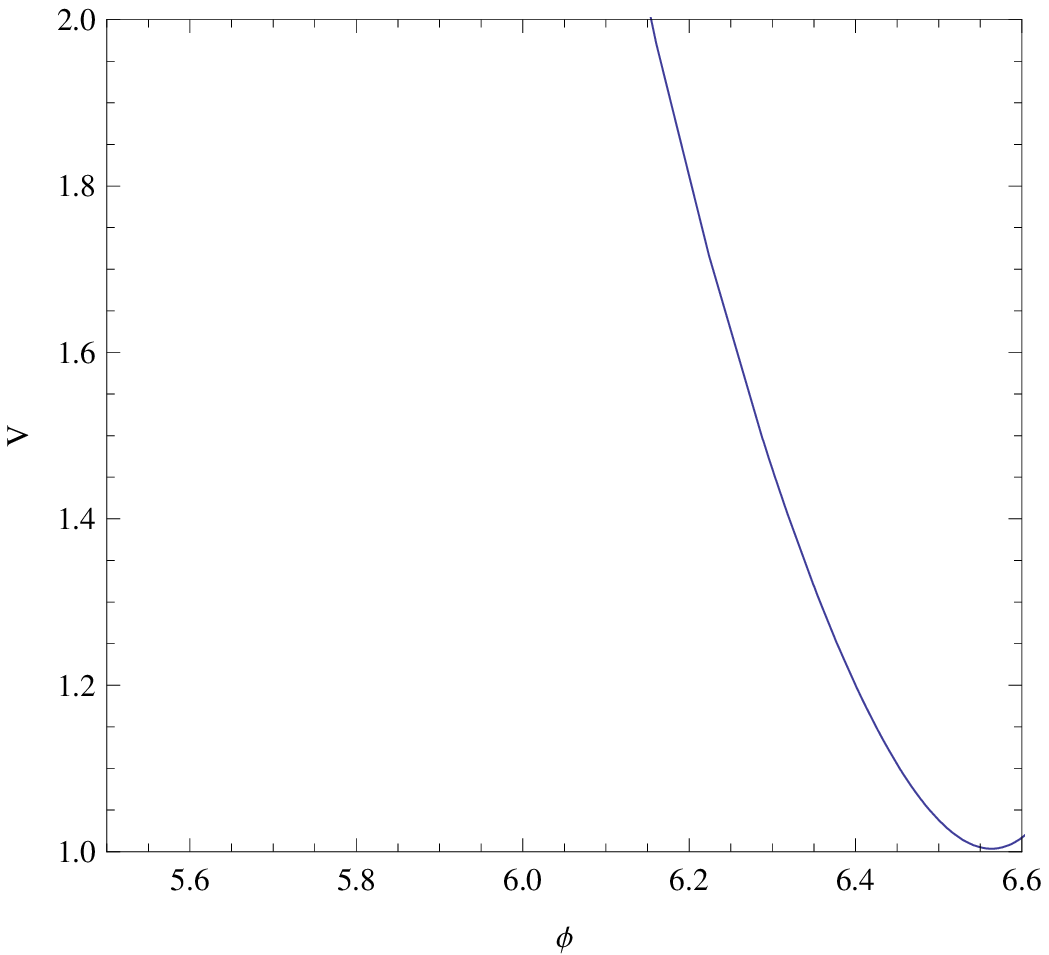}\\
\caption{The variation of $V$ against $\Phi$ for $\sigma=-1$
(phantom field) in VMCG with particular values of parameters
$A_{1}=0.1,B_{0}=0.1,\delta=0.005,\alpha=0.5,C=1,C_{1}=1,n=0.9,m=0.7$.}
\end{figure}
\begin{figure}
\includegraphics[height=2in]{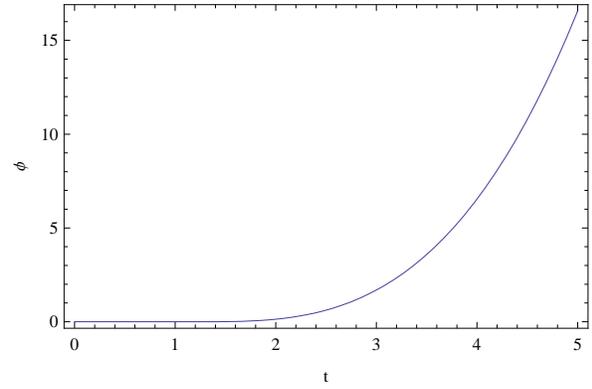}\\
\caption{The variations of $\Phi$ and $t$ against cosmic time $t$
respectively and Fig. 6 represents the variation of $V$ against
$\Phi$ for $\sigma=-1$ (phantom field) in NMCG with particular
values of parameters
$A_{2}=0.1,\beta=0.2,\delta=0.005,\alpha=0.5,C_{1}=1,C_{2}=1,w=-0.9,m=0.7$.}
\end{figure}
\begin{figure}
\includegraphics[height=2in]{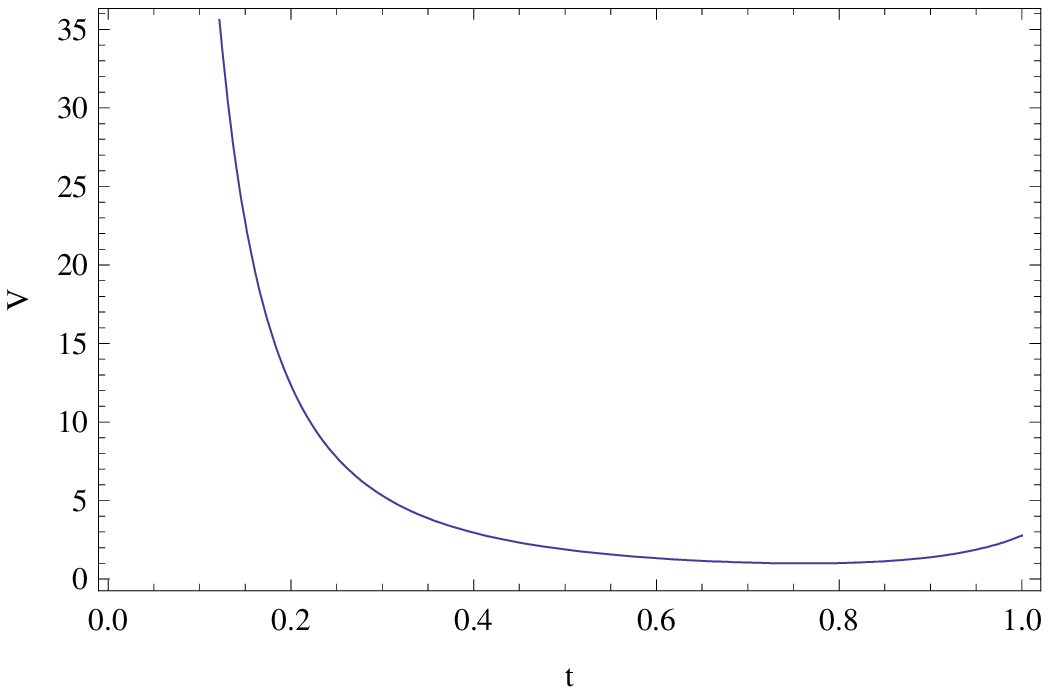}\\
\caption{The variation of $V$ against $t$ for $\sigma=-1$ (phantom
field) in NMCG with particular values of parameters
$A_{2}=0.1,\beta=0.2,\delta=0.005,\alpha=0.5,C_{1}=1,C_{2}=1,w=-0.9,m=0.7$.}
\end{figure}
\begin{figure}
\includegraphics[height=2.5in]{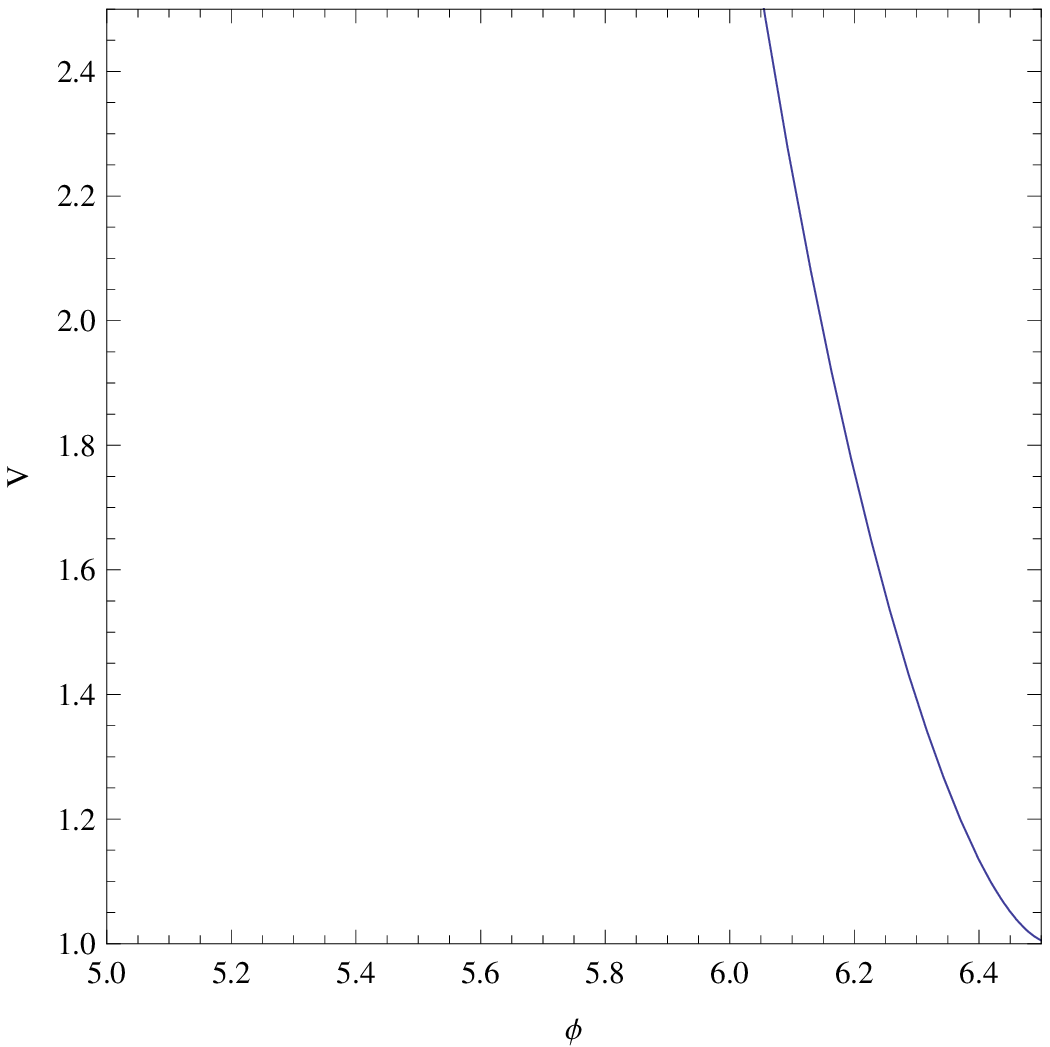}\\
\caption{The variation of $V$ against $\Phi$ for $\sigma=-1$
(phantom field) in NMCG with particular values of parameters
$A_{2}=0.1,\beta=0.2,\delta=0.005,\alpha=0.5,C_{1}=1,C_{2}=1,w=-0.9,m=0.7$.}
\end{figure}

\end{document}